\documentclass[12pt]{article}
\usepackage{a4}
\usepackage{amsmath}
\usepackage{amsfonts}
\usepackage{amsmath}
\usepackage{latexsym}
\usepackage{amsfonts}
\usepackage{graphics}
\usepackage{graphicx}
\usepackage{epic}
\usepackage{eepic}
\usepackage{curves}

\begin{document}
\renewcommand{\thefootnote}{\fnsymbol{footnote}}
\thispagestyle{empty}

\noindent hep-th/   \hfill  December   2004 \\

\noindent \vskip3.3cm
\begin{center}

{\Large\bf The masses of gauge fields in higher spin field theory on
the bulk of $AdS_{4}$}
\bigskip\bigskip\bigskip

{\large Ruben Manvelyan\footnote{On leave from Yerevan Physics
Institute} and Werner R\"uhl}
\medskip

{\small\it Department of Physics\\ Erwin Schr\"odinger Stra\ss e \\
Technical University of Kaiserslautern, Postfach 3049}\\
{\small\it 67653
Kaiserslautern, Germany}\\
\medskip
{\small\tt manvel,ruehl@physik.uni-kl.de}
\end{center}

\bigskip 
\begin{center}
{\sc Abstract}
\end{center}
A local gauge invariant interaction Lagrangian for two gauge fields
of spin $\ell$ and $\ell-2$ $(\ell>2)$ and the scalar field is
defined. It gives rise to one-loop corrections to the gauge field
propagator. The loop function contains the Goldstone boson
propagator for gauge symmetry breaking. The proportionality factor
in front of this propagator is the mass squared of the gauge boson.

\newpage
\section{Introduction}
The $AdS/CFT$ correspondence \cite{Maldacena} opened a new era in
understanding of the strong coupling regime of certain conformal
boundary theories mapping them on the perturbation expansion of the
weakly coupled string/supergravity theory on the bulk $AdS$ space.
In contrast to the most explored case of $AdS_{5}/CFT_{4}$ dealing
with the large $N$ limit of $SU(N)$ $\mathcal{N}=4$ Super Yang-Mills
theory, the $AdS_{4}/CFT_{3}$ correspondence of the critical $O(N)$
sigma model \cite{Klebanov} works in the regime of small t'Hooft
coupling and turns our attention from supersymmetric string/gravity
theories to the theory of massless gauge  higher even spins
\cite{Vasiliev}. This case is interesting also in view of the unique
properties of the renormalization group flow connecting the free
field unstable point of the boundary $O(N)$ vector model with the
stable critical interacting conformal point in the large -$N$ limit
by the deformation with the double trace marginal operator and
Legendre transformation. This flow can be explained from the bulk
side using the same  higher spin theory ($HS(4)$) and different
boundary conditions for the quantized scalar field \cite{WittKleb}.
Note that in the second and nontrivial conformal point of $d=3$
sigma model all higher spin currents except the energy-momentum
tensor (spin two) are conserved only in the large $N$ limit and
their divergence is of first order in $\frac{1}{N}$. On the bulk
side this must correspond to a certain mass generation mechanism on
the one loop level (again order of $\frac{1}{N}$) of the interacting
$HS(4)$ gauge theory.

In this article we explore  this mechanism of the one-loop mass
generation for massless gauge $HS$ fields  from the bulk side
proposed in \cite{Por1} and compare it with the corresponding
boundary sigma model derivation of
 one of the authors \cite{Ruehl}. For this we construct the only one
possible $h^{(\ell)}h^{(\ell-2)}\sigma$ local interaction of fields
with spin $\ell$, $\ell-2$ and zero in the way similar to our
previous investigation of the $h^{(\ell)}\sigma\sigma$ vertex
\cite{Leonhardt,MR}. The coupling constant $g_{\ell}$ of the vertex
$h^{(\ell)}h^{(\ell-2)}\sigma$ remains undetermined, in particular
we do not relate it to the universal coupling constant of $HS(4)$.
Thus the validity of $AdS/CFT$ correspondence on the one-loop level
remains open.
\section{Equations of motion and physical states}
In this article we operate with the following objects: The
conformally coupled scalar field satisfying  the usual equation of
motion in $AdS_{d+1}$ space\footnote{We will use $AdS_{d+1}$
conformal flat metric, curvature and covariant derivatives
commutation rules of the type
\begin{eqnarray}\label{ads}
&&ds^{2}=g_{\mu\nu}dz^{\mu}dz^{\nu}=\frac{L^{2}}{(z^{0})^{2}}\eta_{\mu\nu}dz^{\mu}dz^{\nu},\quad
\eta_{z^{0}z^{0}}=-1, \sqrt{-g}=\frac{1}{(z^{0})^{d+1}}\;,\nonumber\\
&&\left[\nabla_{\mu},\,\nabla_{\nu}\right]V^{\rho}_{\lambda} =
R^{\quad\,\,\rho}_{\mu\nu \sigma}V^{\sigma}_{\lambda}
  -R^{\quad\,\,\sigma}_{\mu\nu \lambda}V^{\rho}_{\sigma}\;,\nonumber\\
 &&R^{\quad\,\,\rho}_{\mu\nu \lambda}=
-\frac{1}{(z^{0})^{2}}\left(\eta_{\mu\lambda}\delta^{\rho}_{\nu} -
\eta_{\nu\lambda}\delta^{\rho}_{\mu}\right)=-\frac{1}{L^{2}}
\left(g_{\mu\lambda}\delta^{\rho}_{\nu}
- g_{\nu\lambda}\delta^{\rho}_{\mu}\right) \;,\nonumber\\
 &&R_{\mu\nu}=-\frac{d}{(z^{0})^{2}}\eta_{\mu\nu}=-\frac{d}{L^{2}}g_{\mu\nu}\quad ,
\quad R=-\frac{d(d+1)}{L^{2}}\;.\nonumber
\end{eqnarray}}
\begin{equation}\label{ss}
    \Box \sigma(z)=\frac{(d+1)(d-1)}{4}\sigma(z) ,
\end{equation}
general spin $\ell $ symmetric and double traceless conserved
currents $J^{(\ell)}_{\mu_{1}\mu_{2}\dots\mu_{\ell}}(z)$, and again
double traceless \cite{Frons} spin $\ell $ gauge fields
$h^{(\ell)}_{\mu_{1}\mu_{2}\dots\mu_{\ell}}(z)$. For shortening the
notation and calculation we contract all symmetric tensors  with the
$\ell$-fold tensor product of a vector $a^{\mu}$. In this notation
Fronsdal's  equation of motion \cite{Frons} for the spin $\ell$
field is
\begin{eqnarray}
  &&{\cal F}(h^{(\ell)}(z;a))=\Box h^{(\ell)}(z;a)-\ell (a\nabla)\tilde{h}^{(\ell-1)}(z;a)
  +\frac{\ell(\ell-1)}{2}(a\nabla)^{2}\bar{h}^{(\ell-2)}(z;a)\quad\\
  &&+\frac{\ell^{2}+\ell(d-5)-2(d-2)}{L^{2}}h^{(\ell)} +
  \frac{\ell(\ell-1)}{L^{2}}a^{2}\bar{h}^{(\ell-2)}(z;a)=0 ,\label{F}\\
  &&\Box_{a}\Box_{a}h^{(\ell)}=0 , \quad \Box_{a}=\frac{\partial^{2}}{\partial
    a^{\mu}\partial a_{\mu}} ,
\end{eqnarray}
where we introduced notations for the trace and divergence
\begin{equation}\label{n}
    \bar{h}^{(\ell-2)}=\frac{1}{\ell(\ell-1)}\Box_{a}h^{(\ell)} ,
    \quad \tilde{h}^{(\ell-1)}=\frac{1}{\ell}\nabla^{\mu}\frac{\partial}{\partial
    a^{\mu}}h^{(\ell)} .
\end{equation}
The basic property of this equation is higher spin gauge invariance
with the traceless parameter $\epsilon^{(\ell-1)}(z;a), $
\begin{equation}\label{gi}
    \delta h^{(\ell)}(z;a)=(a\nabla)\epsilon^{(\ell-1)}(z;a) , \quad
    \Box_{a}\epsilon^{(\ell-1)}(z;a)=0 , \quad \delta {\cal F}(h^{(\ell)}(z;a))=
    0 .
\end{equation}

The equation (\ref{F}) is simplified in the so-called de Donder
gauge
\begin{eqnarray}\label{physeq}
  \tilde{h}^{(\ell-1)}&=& \frac{\ell-1}{2}(a\nabla)\bar{h}^{(\ell-2)} ,\\
  {\cal F}^{dD\,gauge}(h^{(\ell)})&=&\Box h^{(\ell)}
  +\frac{\ell^{2}+\ell(d-5)-2(d-2)}{L^{2}}h^{(\ell)}\nonumber\\&& +
  \frac{\ell(\ell-1)}{L^{2}}a^{2}\bar{h}^{(\ell-2)}=0 .\quad\quad
\end{eqnarray}
It was shown (see for example \cite{Mikh}) that in the de Donder
gauge the residual gauge symmetry leads to the tracelessness of the
on-shell fields. So we can define our massless physical spin $\ell$
modes as traceless and transverse symmetric tensor fields satisfying
the equation (\ref{physeq})
\begin{eqnarray}
  && \left[\Box+\frac{\ell^{2}+\ell(d-5)-2(d-2)}{L^{2}}\right] h^{(\ell)}=0 ,\label{ph1} \\
  &&\bar{h}^{(\ell-2)}=\frac{1}{\ell(\ell-1)}\Box_{a}h^{(\ell)}=0 ,\label{pd2}\\
  &&\tilde{h}^{(\ell-1)}=\frac{1}{\ell}\nabla^{\mu}\frac{\partial}{\partial
    a^{\mu}}h^{(\ell)}=0 .\label{ph3}
\end{eqnarray}
Note that equation (\ref{ph1}) for $\ell=0$ coincides with the
equation for the conformal scalar (\ref{ss}) only for $d=3$.

 In a similar way we can describe the massive higher spin modes using the
following set of constraints on the general symmetric tensor field
$\phi^{(\ell)}(z,a)$ \cite{Waldron}
\begin{eqnarray}
  &&\left[\Box+\frac{\ell^{2}+\ell(d-5)-2(d-2)}{L^{2}}+m^{2}\right] \phi^{(\ell)}=0 , \label{mm1} \\
  &&\bar{\phi}^{(\ell-2)}=\tilde{\phi}^{(\ell-1)}=0 ,
  \label{mm2}\\
  &&L^{2}m^{2}=\Delta(\Delta-d)-(\ell-2)(\ell+d-2) ,\label{mm3}
\end{eqnarray}
where $\Delta$ is the  conformal weight (dimension) of the
corresponding massive (in means of $AdS$ field) representation of
the $SO(d,2)$ isometry group. This general representation
(\ref{mm1})-(\ref{mm3}) with two independent quantum numbers
$[\Delta, \ell]$ under the maximal compact subgroup $SO(d)\times
SO(2)$ goes, after imposing a shortening condition
$\Delta=\ell+d-2$, to the massless higher spin case
(\ref{ph1})-(\ref{ph3}) with the following decomposition
\cite{Ruehl,Por,Por1}
\begin{equation}\label{dec}
    \lim_{\Delta\rightarrow
    \ell+d-2}[\Delta,\ell]=[\ell+d-2,\ell]\oplus[\ell+d-1,\ell-1] .
\end{equation}
The additional massive representation $[\ell+d-1,\ell-1]$ is the
Goldstone field. Reading this decomposition from the opposite side,
we can interpret it as swallowing  of the massive spin $\ell-1$
Goldstone field by the massless spin $\ell$ field with generation of
mass for the latter one. This mass generation can be achieved  by
the following one-loop diagram, \cite{Por1}

\hskip3.3cm \setlength{\unitlength}{0.252mm}
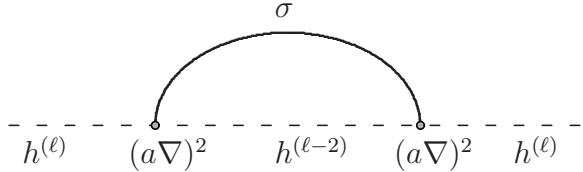
\begin{picture}(341,120)(322,-322)
        \allinethickness{0.252mm}\dashline{6}(322,-273)(623,-273) 
        \allinethickness{0.252mm}\renewcommand{\xscale}{1}\renewcommand{\yscale}{0.7}\put(469,-273){\scaleput(0,0){\Arc(70,-0){180}}} 
        \allinethickness{0.252mm}\special{sh 0.3}\put(399,-273){\ellipse{4}{4}} 
        \allinethickness{0.252mm}\special{sh 0.3}\put(539,-273){\ellipse{4}{4}} 
        \put(329,-293){$h^{(\ell)}$} 
        \put(385,-293){$(a\nabla)^2$} 
        \put(525,-293){\textbf{\textit{$(a\nabla)^2$}}} 
        \put(462,-293){\textbf{\textit{$h^{(\ell-2)}$}}} 
        \put(462,-216){\textbf{\textit{$\sigma$}}} 
        \put(588,-293){\textbf{\textit{$h^{(\ell)}$}}} 
\end{picture}
\noindent where the $h^{(\ell)}h^{(\ell-2)}\sigma$ coupling should
include two derivatives for producing the Goldstone representation
in the transverse traceless part of this loop. This coupling we will
construct in the next section.

\section{The $h^{(\ell)}h^{(\ell-2)}\sigma$ interaction}
In this section we present the interaction Lagrangian responsible
for the one loop mass generation mechanism for HS fields that is
fixed in a unique way by the general concept of \emph{linearized
gauge invariance} (\ref{gi}). We want to construct a gauge invariant
interaction between two double traceless HS fields $h^{(\ell)}(z;a)$
and $h^{(\ell-2)}(z;a)$ and one scalar field $\sigma(z)$. We will
try to realize this interaction in the form of the spin $\ell$ gauge
field $h^{(\ell)}$ times the current $\Psi^{(\ell)}$ constructed
from the spin $\ell-2$ field $h^{(\ell-2)}$, the scalar field
$\sigma$ and \emph{two} ($AdS_{d+1}$) covariant derivatives
$\nabla$.
\begin{eqnarray}
&&S^{(\ell)}_{int}=\frac{1}{\ell}\int
d^{d+1}z\sqrt{g}h^{(\ell)\mu_{1}\dots\mu_{\ell}}\mathcal{J}^{(\ell)}_{\mu_{1}\dots\mu_{\ell}} , \\
  &&
  h^{(\ell)\alpha\beta}_{\alpha\beta\mu_{5}\dots\mu_{\ell}}=0\quad,\quad
  \mathcal{J}^{(\ell)\alpha\beta}_{\alpha\beta\mu_{5}\dots\mu_{\ell}}=0 ,
  \\
  &&\delta_{0}
  h^{(\ell)}_{\mu_{1}\dots\mu_{\ell}}=\partial_{(\mu_{1}}\epsilon_{\mu_{2}\dots\mu_{\ell})},\quad
  \epsilon^{\alpha}_{\alpha\mu_{4}\dots\mu_{\ell}}=0.
\end{eqnarray}
From this point of view the task reduces to the construction of the
current $\mathcal{J}^{(\ell)}(h^{(\ell-2)},\nabla,\nabla,\sigma)$
satisfying the conservation condition
\begin{eqnarray}
   \left[\nabla^{\mu_{1}}\mathcal{J}^{(\ell)}_{\mu_{1}\mu_{2}
   \dots\mu_{\ell}}\right]^{traceless}=0 .\label{cc}
\end{eqnarray}
The conservation condition looks a little bit different from the
usual one due to the double-tracelessness of the gauge field and
current and tracelessness of the corresponding gauge parameter. The
Fronsdal field $\mathcal{J}^{(\ell)}$ can be presented then as
\begin{eqnarray}\label{psi}
&&\mathcal{J}^{(\ell)}(z;a)=J^{(\ell)}(z;a)
+\frac{a^{2}}{2(d+2\ell-3)}\Theta^{(\ell-2)}(z;a) ,\\
&&Tr\mathcal{J}^{(\ell)}(z;a)=\Box_{a}\mathcal{J}^{(\ell)}(z;a)=\Theta^{(\ell-2)}(z;a),\\
&&TrJ^{(\ell)}(z;a)=\Box_{a}J^{(\ell)}(z;a)=0 ,\\
&&Tr\Theta^{(\ell-2)}(z;a)=\Box_{a}\Theta^{(\ell-2)}(z;a)=0 .
\end{eqnarray}
The conservation condition (\ref{cc}) in this representation is
\begin{eqnarray}
 \nabla^{\mu}\frac{\partial}{\partial a^{\mu}} \mathcal{J}^{(\ell)}(z;a)=
 \frac{a^{2}}{2(d+2\ell-5)}Tr\nabla^{\mu}\frac{\partial}
 {\partial a^{\mu}}\mathcal{J}^{(\ell)}(z;a) ,
 \end{eqnarray}
or
\begin{equation}\label{cc1}
\nabla^{\mu}\frac{\partial}{\partial a^{\mu}}
J^{(\ell)}(z;a)+\frac{(a\nabla)
   \Theta^{(\ell-2)}(z;a)}{(d+2\ell-3)} = \frac{a^{2}\nabla^{\mu}\frac{\partial}
   {\partial
   a^{\mu}}\Theta^{(\ell-2)}(z;a)}{(d+2\ell-5)(d+2\ell-3)} .
\end{equation}
For solving this we will introduce the following ansatz for
traceless spin $\ell$ and $\ell-2$ currents
\begin{eqnarray}
&&J^{(\ell)}(z;a) = \sum^{3}_{i=0}A_{i}\left[J^{(\ell)}_{i}(z;a)-
\frac{a^{2}\Box_{a}J^{(\ell)}_{i}(z;a)}{2(d+2\ell
-3)}\right]\nonumber\\
&&\hskip5cm + \frac{a^{2}}{L^{2}}C h^{(\ell-2)}(z)\sigma(z) +
\emph{O}(a^{4},\frac{1}{L^{4}}) ,\label{al}\\
 &&\Theta^{(\ell-2)}(z;a)=
 \sum^{6}_{p=1}B_{p}\left[\Theta^{(\ell-2)}_{p}(z;a)-
\frac{a^{2}\Box_{a}\Theta^{(\ell-2)}_{p}(z;a)}{2(d+2\ell -7)}\right]
+\emph{O}(a^{4},\frac{1}{L^{4}}),\label{bl}
\end{eqnarray}
where we introduced all possible monomials of corresponding order
\begin{eqnarray}
  &&J^{(\ell)}_{1}=(a\nabla)^{2}h^{(\ell-2)}\sigma ,
  \quad J^{(\ell)}_{2}=(a\nabla)h^{(\ell-2)}(a\nabla)\sigma ,
  \quad J^{(\ell)}_{3}=h^{(\ell-2)}(a\nabla)^{2}\sigma ,\\
  && \Theta^{(\ell-2)}_{1}=\nabla_{\mu}h^{(\ell-2)}\nabla^{\mu}\sigma
  ,\quad\Theta^{(\ell-2)}_{2}=h^{(\ell-3)}_{;\mu}(a\nabla)\nabla^{\mu}\sigma
  ,\quad\Theta^{(\ell-2)}_{3}=(a\nabla)h^{(\ell-3)}_{;\mu}\nabla^{\mu}\sigma,\quad\quad\\
  &&\Theta^{(\ell-2)}_{4}=(a\nabla)^{2}\bar{h}^{(\ell-4)}\sigma ,
  \quad\Theta^{(\ell-2)}_{5}=(a\nabla)\bar{h}^{(\ell-4)}(a\nabla)\sigma
 ,\quad\Theta^{(\ell-2)}_{6}=\bar{h}^{(\ell-4)}(a\nabla)^{2}\sigma .
\end{eqnarray}
Substituting this ansatz in the conservation condition (\ref{cc})
and neglecting the noncommutativeness of the covariant derivatives
we will come to  the set of liner equations for the coefficients
$A_{i}$ and $B_{p}$ with the unique solution up to an overall
normalization constant $A$
\begin{eqnarray}
  A_{1} &=& \frac{1}{2}A_{2}=A_{3}=A ,\label{A} \\
  B_{1}&=& -4(d+2\ell-4)A ,\\
  B_{2}&=& B_{3}=-(\ell-2)(d+2\ell-7)A \\
  B_{4}&=& -(\ell-2)(\ell-3)(\frac{1}{2}(d+1)+\ell-5)A \\
  B_{5}&=& -(\ell-2)(\ell-3)(\frac{1}{2}(d+1)+\ell-6)A \\
  B_{6}&=& (\ell-2)(\ell-3)A.
\end{eqnarray}
From (\ref{A}) we see that the leading term of our current
$\mathcal{J}^{(\ell)}$ is the double full derivative
\begin{equation}\label{psiell}
   \mathcal{J}^{(\ell)}(z;a)=A(a\nabla)^{2}\left(h^{(\ell-2)}(z;a)\sigma(z)\right) +
    \text{traces} .
\end{equation}
Now we can restore noncommutativeness of derivatives
\begin{eqnarray}
  &&\Box_{a}(a\nabla)^{2}\left(h^{(\ell-2)}\sigma\right)= 4\Theta^{(\ell-2)}_{1}+
  4(\ell-2)\left(\Theta^{(\ell-2)}_{2}
  +\Theta^{(\ell-2)}_{3}\right)\nonumber\\&&+
   (\ell-2)(\ell-3)\left(3\Theta^{(\ell-2)}_{4}+4\Theta^{(\ell-2)}_{5}+
\Theta^{(\ell-2)}_{6}\right)+\frac{g_{1}(\ell)}{L^{2}}h^{(\ell-2)}\sigma+
\emph{O}(a^{2}),\\
&&\nabla^{\mu}\frac{\partial}{\partial a^{\mu}}
(a\nabla)^{2}\left(h^{(\ell-2)}\sigma\right)
=(a\nabla)\left[4\Theta^{(\ell-2)}_{1}+
(\ell-2)\left(\Theta^{(\ell-2)}_{2}+\Theta^{(\ell-2)}_{3}\right)\right.\nonumber\\
&&\left.+ \frac{(\ell-2)(\ell-3)}{2}\left(\Theta^{(\ell-2)}_{4}
+\Theta^{(\ell-2)}_{5}\right)\right]+\frac{g_{2}(\ell)}{L^{2}}(a\nabla)(h^{(\ell-2)}\sigma)+
\emph{O}(a^{2}) ,
\end{eqnarray}
and calculate the curvature ($\frac{1}{L^{2}}$) correction
coefficient $C$ in (\ref{al})
\begin{eqnarray}
&&C=\frac{1}{2}\left(\frac{g_{1}(\ell)}{d+2\ell-3}-g_{2}(\ell)\right) ,\\
&&g_{1}(\ell)=\frac{(d+1)^{2}}{2}+3d+8\ell-25 , \quad g_{2}(\ell)=
\frac{(d+1)^{2}}{2}+5d+8\ell-33 .
\end{eqnarray}

It is easy to see that if we try to construct a traceless and
conserved  current, we have to set all the coefficients $B_{p}$ to
zero and as a result will get zero for all $A_{i}$. So we deduce
that contrary to the $h^{(\ell)}\sigma\sigma$ case \cite{MR}, the
interaction of the type $h^{(\ell)}h^{(\ell-2)}\sigma$ exists only
in the Fronsdal formulation with a double traceless current. At the
end of this section it is worth to note that our interaction is
unique because it is fixed by gauge invariance. Of course we
considered only invariance with respect to transformation of only
the highest spin $\ell$ gauge field participating in the
interaction, but we can assume on this level of consideration that
variation of the scalar and spin $\ell-2$ fields containing more
derivatives of the parameter $\epsilon$ (see \cite{MR}) will be
compensated by variations of other unknown terms of the interacting
Lagrangian.

\section{Massive higher spin states and Goldstone boson}
Here we consider the origin of the Goldstone boson for general
massless spin $\ell$ field. We can present this mechanism in two
ways generalizing the consideration for the graviton of Ref.
\cite{Por}. First of all we can try to extract the Goldstone field
from the longitudinal part of the spin $\ell$ physical mode
represented by the set of equations (\ref{ph1})-(\ref{ph3}). For
this purpose we insert in the latter set the following ansatz
\begin{equation}\label{ans}
    h^{(\ell)}(z;a)=(a\nabla)\phi^{(\ell-1)}(z;a) .
\end{equation}
Then using the relation:
\begin{eqnarray}
  &&[\nabla_{\mu},(a\nabla)]\partial_{a_{\mu}}\phi^{(\ell-1)}
  =\frac{(\ell-1)(\ell+d-2)}{L^{2}}
  \phi^{(\ell-1)}-\frac{(\ell-1)(\ell-2)a^{2}}{L^{2}}\bar{\phi}^{(\ell-3)}\quad \label{com1}
\end{eqnarray}
we obtain the following equations for $\phi^{(\ell-1)}(z;a)$
\begin{eqnarray}
  && \left[\Box+\frac{(\ell-1)(\ell+d-2)}{L^{2}}\right] \phi^{(\ell-1)}=0 ,\label{gph1} \\
  &&\bar{\phi}^{(\ell-3)}=\frac{1}{\ell(\ell-1)}\Box_{a}\phi^{(\ell)}=0 ,\label{gph2}\\
  &&\tilde{\phi}^{(\ell-2)}=\frac{1}{\ell}\nabla^{\mu}\frac{\partial}{\partial
    a^{\mu}}\phi^{(\ell-1)}=0 .\label{gph3}
\end{eqnarray}
Comparing with (\ref{mm1}) and (\ref{mm3}) we deduce that our field
$\phi^{(\ell-1)}(z;a)$ is the massive spin $\ell-1$ representation
of the isometry group with $\Delta=\ell+d-1$ and should describe the
corresponding Goldstone mode.

The important point of this consideration is the following: For the
traceless spin $\ell-1$ field $\phi^{(\ell-1)}(a;z)$ the correct
second order equation (\ref{gph1}) for the massive states with
$\Delta=\ell+d-1$ originated not only from the gauge fixed equation
of motion (\ref{ph1}) for the massless field
$h^{(\ell)}=(a\nabla)\phi^{(\ell-1)}$ but also from the gauge
condition (\ref{ph3}) using only the commutation rule (\ref{com1}).
In other words we see that
\begin{eqnarray}
  && \nabla^{\mu}\frac{\partial}{\partial a^{\mu}}(a\nabla)
  \phi^{(\ell-1)}=K_{G}^{-1}\phi^{(\ell-1)} , \label{gprop}
\end{eqnarray}
where $K_{G}^{-1}=\Box+\frac{(\ell-1)(\ell+d-2)}{L^{2}}$ is the
inverse propagator for the Goldstone mode $[\ell+d-1,\ell-1]$. This
allows us to formulate a field theoretical realization of the
representation decomposition formula (\ref{dec}) in a similar way as
it was done in \cite{Por} for the graviton case.

Let us describe the massive spin $\ell$ representation
$[\Delta,\ell]$ by the gauge invariant massless action $S_{{\cal
F}}^{GI}[h^{(\ell)}]$, leading to the Fronsdal equation (\ref{F}),
perturbed by the mass term breaking gauge invariance. Neglecting the
traces we can write this mass term in the usual way
\begin{eqnarray}
  &&S_{m}[h^{(\ell)}]= \frac{m^{2}}{2}\int d^{d+1}z
\sqrt{-g}h^{(\ell)\mu_{1}\dots\mu_{\ell}}h^{(\ell)}_{\mu_{1}\dots\mu_{\ell}} ,
\label{massterm}\\
  && L^{2}m^{2}=\Delta(\Delta-d)-(\ell-2)(\ell+d-2) .
\end{eqnarray}
The action $S_{{\cal F}}^{GI}[h^{(\ell)}]+S_{m}[h^{(\ell)}]$ is not
gauge invariant and corresponds to the left hand side of the
relation (\ref{dec}). To get a description for the right hand side
of (\ref{dec}) we have to restore gauge invariance for the spin
$\ell$ field introducing the Goldstone spin $\ell-1$ field by
shifting $h^{(\ell)}$ with $h^{(\ell)}+(a\nabla)\phi^{(\ell-1)}$.
This will affect only the mass term (\ref{massterm})
\begin{eqnarray}
  && S_{m}[h^{(\ell)},\phi^{\ell-1}]= \frac{m^{2}}{2}\int d^{d+1}z
\sqrt{-g}(h^{(\ell)}_{\mu_{1}\dots\mu_{\ell}}+
\nabla_{(\mu_{1}}\phi^{(\ell-1)}_{\mu_{2}\dots\mu_{\ell})})^{2}
.\label{massterm1}
\end{eqnarray}
The St\"uckelberg action $S_{{\cal
F}}^{GI}[h^{(\ell)}]+S_{m}[h^{(\ell)},\phi^{\ell-1}]$ is gauge
invariant under the gauge transformations (\ref{gi}) and
$\delta\phi^{(\ell-1)}(a;z)=-\epsilon^{(\ell-1)}(a;z)$ and describes
the right hand side of (\ref{dec}). Moreover after integrating out
the Goldstone field $\phi^{(\ell-1)}$ in (\ref{massterm1}) we obtain
according to the relation (\ref{gprop}) the following part of the
effective action
\begin{eqnarray}
  && \frac{m^{2}}{2}\int d^{d+1}z_{1}
\sqrt{-g(z_{1})}\int d^{d+1}z_{2} \sqrt{-g(z_{2})}\nabla\cdot
h^{(\ell)}(z_{1})K_{G}(z_{1};z_{2})\nabla\cdot
h^{(\ell)}(z_{2})\quad\quad\label{mterm}\nonumber\\
&&\qquad\qquad\qquad\qquad\qquad + \textnormal{trace terms}.
\end{eqnarray}
This dramatically simplifies  the evaluation of the mass of the
higher spin field generated by the one loop graph of section 1 with
the $h^{(\ell)}h^{(\ell-2)}\sigma$ interaction constructed in the
previous section: It is enough  to evaluate in the corresponding
loop function the coefficient in front of the term with the
behaviour of the Goldstone mode propagator between two divergences
of the external $h^{(\ell)}$ fields (\ref{mterm}). This will be done
in the next section.

Another way to find the Goldstone mode is to construct directly
transverse traceless states of $h^{(\ell)}(z;a)$. For doing that we
introduce first so-called Lichnerowicz operator \cite{Lichn}. In
$AdS_{d+1}$ for a general rank $\ell$ symmetric tensor this operator
looks like
\begin{equation}\label{Lichn}
    \mathbb{L}_{(\ell)}h^{(\ell)}(z;a)=\Box h^{(\ell)}(z;a)
    -\frac{\ell(\ell+d-1)}{L^{2}}h^{(\ell)}(z;a)
    +\frac{\ell(\ell-1)a^{2}}{L^{2}}\bar{h}^{(\ell-2)}(z;a) ,
\end{equation}
and obeys the following conditions
\begin{eqnarray}
  &&\frac{1}{\ell(\ell-1)}\Box_{a} \mathbb{L}_{(\ell)}h^{(\ell)}(z;a)
  =\mathbb{L}_{(\ell-2)}\bar{h}^{(\ell-2)}(z;a) ,\label{trd1} \\
  &&\frac{1}{\ell}\nabla^{\mu}\partial_{a^{\mu}}\mathbb{L}_{(\ell)}h^{(\ell)}(z;a)
  =\mathbb{L}_{(\ell-1)}\bar{h}^{(\ell-1)}(z;a) ,\label{trd2} \\
  &&(a\nabla)\mathbb{L}_{(\ell)}h^{(\ell)}(z;a)=
  \mathbb{L}_{(\ell+1)}(a\nabla)h^{(\ell)}(z;a) .\label{trd3}
\end{eqnarray}
This commutativeness with the covariant derivatives and trace
operation allows us to drop the degrees index of
$\mathbb{L}_{(\ell)}$ and consider it as a number in the calculation
of the transverse traceless projection of $h^{(\ell)}(z;a)$. We can
expand this projection in the following series
\begin{eqnarray}
  &&h^{(\ell)tt}(z;a)=h^{(\ell)}+A(\mathbb{L})(a\nabla)\tilde{h}^{(\ell-1)}
  +B(\mathbb{L})(a\nabla)^{2}\tilde{\tilde{h}}^{(\ell-2)}\nonumber\\
  &&+C(\mathbb{L})(a\nabla)^{2}\bar{h}^{(\ell-2)} +
   \textsl{O}(a^2,(a\nabla)^3)+\dots .\label{proj}
\end{eqnarray}
The tracelessness condition will express the next coefficients in
terms of the previous ones but for us more interesting is the
transversity condition
$\nabla^{\mu}\partial_{a^{\mu}}h^{(\ell)tt}=0$   leading to the
following solution for the coefficient $A(\mathbb{L})$
\begin{equation}\label{A1}
    A(\mathbb{L})=\frac{-\ell}{\Box+\frac{(\ell-1)(\ell+d-2)}{L^{2}}}=
    \frac{-\ell}{\mathbb{L}^{t}_{(\ell-1)}+\frac{2(\ell-1)(\ell+d-2)}{L^{2}}}
    ,
\end{equation}
here $\mathbb{L}^{t}_{(\ell-1)}$ is the Lichnerowicz operator
(\ref{Lichn}) for the rank $\ell-1$ traceless states. We see that
the first pole in the $tt$ projection behaves exactly as a
propagator for a Goldstone tensor boson.

\section{Propagator and loop graph}
We study a perturbative expansion of the gauge field propagator

\hskip-0.8cm\setlength{\unitlength}{0.252mm}
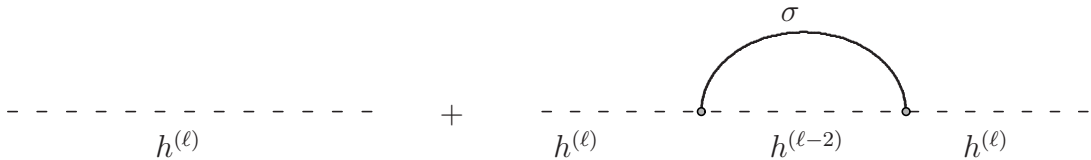
\begin{picture}(571,104)(96,-156)
        \allinethickness{0.252mm}\dashline{6}(96,-114)(288,-114) 
        \allinethickness{0.252mm}\dashline{6}(378,-114)(666,-114) 
        \allinethickness{0.252mm}\renewcommand{\xscale}{1}\renewcommand{\yscale}{0.7777778}\put(516,-114){\scaleput(0,0){\Arc(54,-0){180}}} 
        \put(174,-138){$h^{(\ell)}$} 
        \put(384,-138){$h^{(\ell)}$} 
        \put(600,-138){$h^{(\ell)}$} 
        \put(498,-138){$h^{(\ell-2)}$} 
        \put(504,-66){$\sigma$} 
        \put(324,-120){$+$} 
        \allinethickness{0.252mm}\special{sh 0.3}\put(462,-114){\ellipse{4}{4}} 
        \allinethickness{0.252mm}\special{sh 0.3}\put(570,-114){\ellipse{4}{4}} 
\end{picture}

\noindent with the derivative action studied in section 3. To
express these graphs we need the propagators for the fields
$h^{(\ell)}, h^{(\ell-2)}$, and $\sigma$. We use the results
presented in \cite{LMR1,LMR2,RM2}. The symmetric traceless tensor
field $h^{(\ell)}$ of rank $\ell$ and dimension $\Delta$ in
$AdS_{d+1}$ is massive in general according to the formula
(\ref{mm3}). This mass vanishes if
\begin{equation}\label{zm}
    \Delta=d+\ell-2 .
\end{equation}
In this case $h^{(\ell)}$ allows gauge transformations as a
symmetry. Representations $[\Delta, \ell]$ of the universal covering
group of the conformal group $SO(d+1,1)$ are unitary. From this
point on we will use the Euclidian version of the $AdS_{d+1}$ spaces
and set to 1 the $AdS$ radius $L$. Representations satisfying
(\ref{zm}) are also called "exceptional". The propagator of
$h^{(\ell)}$ between two points $z_{1}, z_{2}$ of $AdS_{d+1}$
(symmetric traceless part) is a bitensor, it can be spanned by means
of an algebraic bitensor basis ${I_{1},I_{2},I_{3},I_{4}}$
\cite{LMR1} by
\begin{equation}\label{prop1}
  \langle h^{(\ell)}(z_{1}) h^{(\ell)}(z_{2})\rangle_{AdS}=
  \sum_{\substack{n_{1},n_{2},n_{3},n_{4}\\
  n_{1}+n_{2}+2n_{3}+2n_{4}=\ell}}I_{1}^{n_{1}}I_{2}^{n_{2}}I_{3}^{n_{3}}I_{4}^{n_{4}}
  F^{[\Delta,\ell]}_{n_{1}n_{2}n_{3}n_{4}}(\zeta(z_{1},z_{2})) .
\end{equation}
Here $\zeta$ is the non-Euclidian cosine of the angle between
$z_{1}$ and $z_{2}$
\begin{equation}\label{zeta}
    \zeta(z_{1},z_{2})=\cosh\Phi(z_{1},z_{2})=\frac{(z^{0}_{1})^{2}+(z^{0}_{2})^{2}+(
    \vec{z}_{1}-\vec{z}_{2})^{2}}{2z_{1}^{0}z_{2}^{0}}
\end{equation}
in Poincar\'{e} coordinates.

In the case of gauge fields $h^{(\ell)}$ the propagators are most
elegantly derived in the de Donder gauge by a recursive algorithm
\cite{RM2}. We neglect the trace terms, they can always be added at
the end with simple arguments (see \cite{LMR1}). In order to derive
(\ref{prop1}) we start from
\begin{equation}\label{PS}
    \Psi^{(\ell)}[F]=\sum^{\ell}_{k=0}I^{\ell-k}_{1}I^{k}_{2}F_{k}(\zeta)
\end{equation}
and impose the field equation and the constraints of double
tracelessness and de Donder gauge. We obtain \cite{RM2}
\begin{equation}\label{f0}
F_{0}=C\zeta^{-\Delta}{}_{2}F_{1}(\frac{1}{2}\Delta,\frac{1}{2}
(\Delta+1);\Delta-\mu+1;\zeta^{-2}) ,
\end{equation}
with $\Delta$ (\ref{zm}) and $\mu=\frac{1}{2}d$,
\begin{eqnarray}\label{l1}
  && F_{1}(\zeta)=-\frac{\ell}{\zeta}F_{0}(\zeta) , \\
  &&F_{k}(\zeta)=(-1)^{k}\binom{\ell}{k}\zeta^{-k}F_{0}(\zeta)+f_{k}(\zeta),\quad
  (k\geq 2) ,\label{lk}
\end{eqnarray}
where $f_{k}(\zeta)$ is asymptotically small
\begin{equation}\label{fks}
    \frac{f_{k}(\zeta)}{F_{k}(\zeta)}=O(\frac{1}{\zeta}) ,\quad
    \zeta \rightarrow \infty .
\end{equation}
Inserting these expressions in (\ref{PS}) we obtain
\begin{equation}\label{PSF}
    \Psi^{(\ell)}[F]=\left(I_{1}-\frac{1}{\zeta}I_{2}\right)^{\ell}F_{0}(\zeta)
    +\sum^{\ell}_{k=2}I^{\ell-k}_{1}I^{k}_{2}f_{k}(\zeta) .
\end{equation}
In the limit $\zeta\rightarrow \infty$ Dobrev's bulk-to-boundary
propagator \cite{Dobrev} can be derived from (\ref{PSF}) (except the
trace terms).

We set the normalization constant $C$ equal to
\begin{equation}\label{C}
    C=\frac{\Gamma(\frac{1}{2}\Delta)\Gamma(\frac{1}{2}(\Delta+1))}
    {(4\pi)^{\mu+\frac{1}{2}}\Gamma(\Delta-\mu+1)} .
\end{equation}
 Using the fact that for $\zeta\rightarrow 1$
\begin{eqnarray}
  &&{}_{2}F_{1}(\frac{1}{2}\Delta,\frac{1}{2}(\Delta+1);\Delta-\mu+1;\zeta^{-2})\nonumber \\
  &&= \frac{\Gamma(\Delta-\mu+1)\Gamma(\mu-\frac{1}{2})}{\Gamma(\frac{1}{2}\Delta)
  \Gamma(\frac{1}{2}(\Delta+1))}(\zeta^{2}-1)^{-\mu+\frac{1}{2}}+O(1)
   , \quad \Re\mu>\frac{1}{2} , \label{glim}
\end{eqnarray}
and
\begin{equation}
\Box\frac{\Gamma(\mu-\frac{1}{2})}{(4\pi)^{\mu+\frac{1}{2}}}(\zeta^{2}-1)^{-\mu+\frac{1}{2}}
  =-\delta(z_{1},z_{2})+\text{regular terms} ,\label{delta}
  \end{equation}
 we see that $F_{0}(\zeta)$ appears as the kernel for the inverse wave
operator $(-\Box + m^{2})$ for the massive scalar field in Euclidian
$AdS_{d+1}$ space with  $m^{2}=\Delta(\Delta-d)$. In the next
section we shall use this normalization (the same was used by
D'Hoker, Freedman and coauthors \cite{Freed} for the vector and
graviton) for all symmetric traceless tensor fields be they massless
or not.

Summarizing the consideration of this section we can formulate the
following statements

a)The double traceless spin $\ell$ Fronsdal  propagator in the same
fashion as the corresponding double traceless spin $\ell$ current
(see e.g. \cite{MR} and section 2) can be expressed as a combination
of two traceless spin $\ell$ and $\ell-2$ propagators. The spin
$\ell-2$ propagator is connected with the Goldstone boson
\cite{RM2}.

b)The leading term which is  useful for our calculation in the next
section, can be expanded as in (\ref{PS}) with the first and most
important contribution $F_{0}$ satisfying the massive scalar
equation
\begin{equation}\label{masscal}
    [-\Box+\Delta(\Delta-d)]F_{0}(\zeta(z_{1},z_{2}))=-\delta(z_{1},z_{2})
    ,
\end{equation}
where $\Delta$ is the dimension of the representation
$[\Delta,\ell]$ described by the initial gauge or massive field
$h^{(\ell)}$.

The precise proof of the propagator structure is presented in
\cite{RM2}.

\section{The loop function}

In the coordinate space the loop function is the product of the
propagators for the fields $h^{(\ell-2)}$ and $\sigma$ (from now on
$d=3$)
\begin{equation}\label{sigmaprop}
    \langle\sigma(z_{1})\sigma(z_{2})\rangle_{AdS_{4}}
    =\frac{1}{(4\pi\zeta)^{2}}K(\zeta), \quad K(\zeta)=\frac{\zeta^{2}}{\zeta^{2}-1}
\end{equation}
We present the propagator of $h^{(\ell-2)}$ by its $F_{0}$ term and
introduce in (\ref{psiell}) a coupling constant which is yet unknown
\begin{equation}\label{g}
    A=\frac{g_{\ell}}{\sqrt{N}} .
\end{equation}
Then the leading term of the loop function is
\begin{eqnarray}
  &&\frac{g_{\ell}^{2}}{N}\frac{\Gamma(\frac{1}{2}(\ell-1))\Gamma(\frac{1}{2}\ell)}
  {\Gamma(\ell-\frac{3}{2})(4\pi)^4}I^{\ell-2}_{1}\zeta^{-(\ell+1)}
  K(\zeta){}_{2}F_{1}\left[\frac{1}{2}(\ell-1),\frac{1}{2}
  \ell;\ell-\frac{3}{2};\zeta^{-2}\right] .\label{loopf}
 \end{eqnarray}
Note that in $AdS$ space the product of two scalar propagators is
not one scalar propagator but an infinite sum of scalar propagators
with dimensions increasing
\begin{equation}\label{dincr}
\Delta_{(n)}=\Delta_{1}+\Delta_{2}+2n ,\quad n \in \mathbb{N}_{0} .
\end{equation}
As was described in the previous sections we have to use our
interaction (\ref{psiell}) and calculate the contribution of this
loop function to the propagation of the longitudinal modes of
$h^{(\ell)}$. Actually we have to extract the term $\nabla\cdot
h^{(\ell)}G_{[\ell+2,\ell-1]}\nabla\cdot h^{(\ell)}$, where
$G_{[\ell+2,\ell-1]}$  is the Goldstone mode propagator
(\ref{gprop}) \cite{Por,Por1} normalized in a proper way. For this
we have to take the gradients of (\ref{loopf}) in the following
sense: For a bitensor $\Psi^{(\ell)}$ such as (\ref{PS}) we
introduce a map
\begin{equation}\label{map}
    \Psi^{(\ell)}\rightarrow
    \Psi^{(\ell+1)}=(a\cdot\nabla_{1})(c\cdot\nabla_{2})\Psi^{(\ell)}(\zeta;a,c)=
    \sum^{\ell+1}_{n=0}I^{\ell+1-n}_{1}I^{n}_{2}G_{n}(\zeta) +
    \text{trace terms} ,
\end{equation}
with
\begin{equation}\label{gf}
    G_{n}=F''_{n-1}(\zeta)+(2n+1)F'_{n}(\zeta)+(n+1)^{2}F_{n+1} .
\end{equation}
For $n=0$ this gives
\begin{equation}\label{G0}
    G_{0}(\zeta)=F'_{0}(\zeta)+F_{1}(\zeta) .
\end{equation}
$F_{1}$ is related to $F_{0}$ by (\ref{l1})
\begin{equation}\label{f1}
    F_{1}=-\frac{\ell-2}{\zeta}F_{0}.
\end{equation}
 Applying the gradient transformation to
(\ref{loopf}) we obtain then from (\ref{G0})
\begin{equation}\label{gl}
-\frac{g^{2}_{\ell}}{N}\frac{\Gamma(\frac{1}{2}(\ell-1))
\Gamma(\frac{1}{2}\ell)}{\Gamma(\ell-\frac{3}{2})(4\pi)^4}(2\ell-1)I^{\ell-1}_{1}
\zeta^{-(\ell+2)}\left(1+O(\frac{1}{\zeta^{2}})\right) .
\end{equation}
Next we have to identify in this expression the normalized Goldstone
boson propagator of the representation $[\ell+2,\ell-1]$. For this
purpose we use a normalizing factor analogous to (\ref{C}) and get
\begin{equation}\label{nf}
    \frac{\Gamma(\frac{1}{2}(\ell+2))\Gamma(\frac{1}{2}(\ell+3))}
    {\Gamma(\ell+\frac{3}{2})(4\pi)^{2}}I^{\ell-1}_{1}\zeta^{-(\ell+2)}{}_{2}F_{1}
    \left[\frac{1}{2}(\ell+2),\frac{1}{2}
  (\ell+3);\ell+\frac{3}{2};\zeta^{-2}\right] .
\end{equation}
Dividing (\ref{gl}) by (\ref{nf}) we obtain the mass squared of
$h^{(\ell)}$
\begin{eqnarray}
m^{2}_{\ell}&=&\frac{g^{2}_{\ell}}{N}\frac{1}{(4\pi)^{2}}
\frac{(2\ell-1)\Gamma(\frac{1}{2}(\ell-1))
\Gamma(\frac{1}{2}\ell)\Gamma(\ell+\frac{3}{2})}
{\Gamma(\frac{1}{2}(\ell+2))\Gamma(\frac{1}{2}(\ell+3))\Gamma(\ell-\frac{3}{2})}\nonumber\\
&=&\frac{g^{2}_{\ell}}{N}\frac{1}{(\pi)^{2}}
\frac{(\ell-\frac{3}{2})(\ell-\frac{1}{2})^{2}(\ell+\frac{1}{2})}{(\ell-1)
\ell(\ell+1)} .\label{ml}
\end{eqnarray}
Remembering that the whole mechanism works only for $\ell\geq 3$ we
should multiply (\ref{ml}) with $1-\delta_{\ell2}$.

Comparing with the results obtained from the $O(N)$ sigma model and
$AdS/CFT$ correspondence \cite{Ruehl}
\begin{equation}\label{m3}
    m^{2}_{\ell}=\frac{16}{3N\pi^{2}}(\ell-2)+ O(\frac{1}{N^{2}}) ,
\end{equation}
we can obtain the interaction constant $g_{\ell}$
\begin{equation}\label{gell}
    g^{2}_{\ell}=\frac{16}{3}\frac{(\ell-2)_{4}}
    {(\ell-\frac{1}{2})(\ell-\frac{3}{2})_{3}} .
\end{equation}

An independent derivation of this coupling constant will be
presented elsewhere.

\subsection*{Acknowledgements}
\quad R.M. thanks W. Nahm for valuable discussions and Dublin IAS
for hospitality. The work of R.~M. was supported by DFG (Deutsche
Forschungsgemeinschaft) and in part by the INTAS grant \#03-51-6346.

\end{document}